\documentclass[11pt]{article} 
\usepackage{epsfig,amssymb} 
\input{epsf} 
\textwidth17.cm \newcommand{\beq}{\begin{equation}} 
\newcommand{\eeq}{\end{equation}} 
\newcommand{\beqa}{\begin{eqnarray}} 
\newcommand{\eeqa}{\end{eqnarray}}

\newcommand{\vs}{\vspace{-0.20cm}} 

\begin{document} 


\begin{flushright} 
{\small  HISKP-TH-03/21}
\end{flushright} 

\vspace{1in}

\begin{center} 

\bigskip 

{\Large\bf On the pion cloud of the nucleon} 

\end{center} 

\vspace{.3in} 

\begin{center} 

{\large H.-W. Hammer$^{\star,}$\footnote{email: 
hammer@itkp.uni-bonn.de}$^,$\footnote{Address after Jan. 1$^{st}$, 2004: 
Institute for Nuclear Theory, University of Washington, Seattle,  
WA 98195-1550, USA.}, 
D. Drechsel$^{\dagger,}$\footnote{email: drechsel@kph.uni-mainz.de}, 
and Ulf-G. Mei{\ss}ner$^{\star,\ddagger,}$\footnote{email: 
meissner@itkp.uni-bonn.de} } 

\vspace{1cm} 

$^\star${\it Universit\"at Bonn, Helmholtz--Institut f\"ur Strahlen-- und 
Kernphysik (Theorie)\\ Nu{\ss}allee 14-16, D-53115 Bonn, Germany} 

\bigskip 
$^\dagger${\it Universit\"at Mainz, Institut f\"ur Kernphysik, 
J-J.-Becher Weg 45, D-55099 Mainz, Germany} 

\bigskip 
$^\ddagger${\it Forschungszentrum J\"ulich, Institut f\"ur Kernphysik 
(Theorie)\\ D-52425 J\"ulich, Germany} 

\bigskip 

\bigskip 

\end{center} 

\vspace{.6in} 

\thispagestyle{empty} 

\begin{abstract}
\noindent 
We evaluate the two--pion contribution to the nucleon electromagnetic form 
factors by use of dispersion analysis and chiral perturbation theory. After 
subtraction of the rho--meson component, we calculate the distributions 
of charge and magnetization in coordinate space, which can be interpreted as 
the effects of the pion cloud. We find that the charge distribution of this 
pion cloud effect peaks at distances of about 0.3 fm. Furthermore, we
calculate the contribution of the pion cloud to the isovector charges and 
radii  of the nucleon.
\end{abstract} 

\vfill 

\pagebreak 

\noindent 
{\bf 1.} That the pion plays an important role in the long--range structure 
of the nucleon is known since long. This can for example be deduced from the 
phenomenological analysis of the nucleon electromagnetic form factors (for an 
early attempt within meson theory see e.g. \cite{FHK}). However, only with the 
advent of QCD and its spontaneously broken chiral symmetry, in which the pions 
emerge as pseudo--Goldstone bosons, this concept could be put on a firmer 
basis. Exploiting the chiral symmetry of QCD, the long-range low--momentum 
structure of the nucleon can be calculated within chiral perturbation theory 
(CHPT), which is the low-energy effective field theory of the Standard
Model. For calculations of these form factors within CHPT, see 
\cite{GSS}-\cite{Mz}. Furthermore, since vector mesons play an important 
role in the electromagnetic structure of the nucleon, see e.g. 
\cite{FF}-\cite{Dubni}, 
care must be taken  in attributing a certain size or length scale to the 
various contributions (this is discussed in some detail in section 2 of 
\cite{BHMcut}). A new twist to this picture was given in the recent paper of 
Ref.~\cite{FW}, where an interpretation of the form factor data was given in 
terms of a phenomenological fit with an ansatz for the pion cloud based on the 
old idea that the proton can be thought of as virtual neutron-positively 
charged pion pair. A very long--range contribution to the charge distribution 
in the Breit--frame extending out to about 2~fm was found and attributed to 
the pion cloud. While naively the pion Compton wave length is of this size, 
these findings are indeed surprising if compared with the ``pion cloud'' 
contribution due to the two--pion contribution for the isovector spectral 
functions of the nucleon form factors, which can be obtained from unitarity or 
chiral perturbation theory. As it will turn out these latter contributions to 
the long--range part of the nucleon structure are much more confined in space 
and agree well with earlier (but less systematic) calculations based on chiral 
soliton models, see e.g. \cite{UGM}. Therefore it remains to be shown how to 
reconcile the findings of Ref.~\cite{FW}, based on a global fit to all nucleon 
form factors, with the results of dispersion analysis and chiral perturbation 
theory. 

\medskip 

\noindent 
{\bf 2.} First, we must collect some basic definitions. The nucleon 
electromagnetic form factors are defined by the nucleon matrix element of the 
quark electromagnetic current, 
\beq 
\langle N(p') | \bar q \gamma^\mu {\mathcal Q} q | N(p)\rangle 
= \bar u (p') \left[ \gamma^\mu \, F_1 (q^2) + \frac{i}{2m} 
\sigma^{\mu\nu} (p'-p)_\nu \,  F_2 (q^2) \right] u(p)~, 
\eeq 
with $q^2 = (p'-p)^2 =t$ the invariant momentum transfer squared, 
${\mathcal Q}$ the quark charge matrix, and $m$ the nucleon mass. $F_1 (q^2)$ 
and $ F_2 (q^2)$ are the Dirac and the Pauli form factors, respectively. 
Following the conventions of \cite{MMD}, we decompose the form factors into 
isoscalar ($S$) and isovector ($V$) components, 
\beq 
F_i (q^2) =  F_i^S (q^2) + \tau_3 \,  F_i^V (q^2)~,\quad i=1,2\,, 
\eeq 
subject to the normalization 
\beq F_1^S (0) = F_1^V (0) = \frac{1}{2}~, \quad F_2^{S,V} (0) 
= \frac{\kappa_p \pm \kappa_n}{2}~, 
\eeq 
with $\kappa_p \, (\kappa_n) = 1.793\, (-1.913)$ the anomalous magnetic moment 
of the proton (neutron). We will also
use the Sachs form factors, 
\beq
\label{conv}
G_E^I (q^2) = F_1^I (q^2) + \frac{q^2}{4m^2} F_2^I (q^2)~, 
\quad G_M^I (q^2) = F_1^I (q^2) + F_2^I (q^2)~, \quad I = S, V ~. 
\eeq 
These are commonly referred to as the electric and the magnetic nucleon form 
factors. The slope of the form factors at $q^2=0$ can be expressed in terms 
of a nucleon radius 
\beq 
\label{radidef} 
\langle r^2\rangle_i^I = \frac{6}{F_i^I (0)}\left.
\frac{dF_i^I (q^2)}{dq^2} \right|_{q^2=0}\,,\quad i=1,2\,, \quad I = S, V\,,
\eeq 
and analogously for the Sachs form factors. The analysis of the  nucleon 
electromagnetic form factors proceeds most
directly through the spectral representation given by\footnote{We work here 
with unsubtracted dispersion relations. Since the normalizations of all the 
form factors are known, one could also work with once--subtracted dispersion 
relations. For the topic studied here, this is of no relevance.} 
\beq\label{spec} F_i^I (q^2) = \frac{1}{\pi}\, \int_{{(\mu_0^I)}^2}^\infty 
\frac{\sigma_i^I (\mu^2) \, d\mu^2}{\mu^2 - q^2}~, \quad i =1,2~, 
\quad I = S, V~, 
\eeq 
in terms of the real spectral functions 
$\sigma_i^I (\mu^2) = {\rm Im}\,F_i^I (\mu^2)$, 
and the corresponding thresholds are given by 
$\mu_0^S = 3M_\pi$, $\mu_0^V = 2M_\pi$. 
Since the isovector spectral function is non--vanishing for smaller momentum 
transfer (starting at the two--pion cut) than the isoscalar one (starting at 
the three--pion cut), we will mostly consider the isovector spectral functions 
in what follows. We consider the nucleon form factors in the space-like region.
In the Breit--frame (where no energy is transferred), any form factor $F$ can 
be written as the Fourier--transform of a coordinate space density, 
\beq 
F({\bf q}^2) = \int d^3{\bf r}\, {\rm e}^{i {\bf q}\cdot {\bf r}}\, 
\rho(r)~, 
\eeq 
with ${\bf q}$ the three--momentum transfer. In particular, comparison with 
Eq.~(\ref{spec}) allows us to express the density $\rho (r)$ in terms of the 
spectral function 
\beq
\label{dens} 
\rho(r) = \frac{1}{4\pi^2}\, \int_{\mu_0^2}^\infty d\mu^2 \, 
\sigma (\mu^2) \, \frac{{\rm e}^{-\mu r}}{r}~. 
\eeq 
Note that for the electric and the magnetic Sachs form factor, $\rho (r)$ is 
nothing but the charge and the magnetization density, respectively. For
the Dirac and Pauli form factors, Eq.~(\ref{dens}) should be considered as
a formal definition. This 
equation expresses the density as a linear combination of Yukawa distributions,
each of mass $\mu$. The lightest mass hadron is the pion, and from 
Eq.~(\ref{dens}) it is evident that pions are responsible for the long--range 
part of the electromagnetic structure of the nucleon. This contribution is 
commonly called the ``pion cloud'' of the nucleon and in fact this long--range 
low--$q^2$ contribution to the nucleon form factors can be directly calculated 
on the basis of chiral perturbation theory, as will be discussed later.

\medskip

\begin{figure}[ht] 
\vspace{0.9cm} 
\centerline{ \epsfysize=1.5in \epsffile{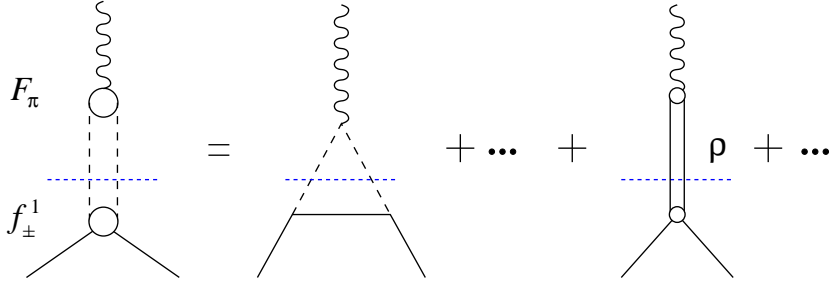} } 
\vspace{0.3cm} 
\begin{center} 
\caption{Two--pion contribution to the isovector nucleon form factors. On the 
left side, the exact representation based on unitarity is shown, whereas the 
triangle diagram on the right side leads to the strong enhancement of the 
isovector spectral functions close to threshold. Also shown is the dominant 
$\rho$--meson contribution. \label{fig:2pic}} \vspace{-0.3cm} 
\end{center} 
\end{figure}

\noindent 
{\bf 3.} Next, we wish to evaluate the two--pion contribution in a 
model--independent way and draw some conclusions on the spatial extent of the 
pion cloud from that. As pointed out long ago~\cite{FF} and further elaborated 
on \cite{HP}, unitarity allows us to determine the isovector spectral 
functions from threshold up to masses of about 1~GeV in terms of the pion 
charge form factor $F_\pi (t)$ and the P--wave $\pi\pi \bar N N$ partial 
waves, see Fig.~\ref{fig:2pic}. We use here the form 
\beqa 
\nonumber 
{\rm Im}~G_E^{V} (t) &=& \frac{q_t^3}{m\sqrt{t}}\, 
\left|F_\pi (t)\right|^2 \, J_+ (t)~,\\ 
{\rm Im}~G_M^{V} (t) &=& \frac{q_t^3}{\sqrt{2t}}\, 
\left|F_\pi (t)\right|^2 \, J_- (t)~, 
\label{uni} 
\eeqa 
where $q_t=\sqrt{t/4-M_\pi^2}$. The functions $J_{\pm} (t)$ are related to the 
$t$--channel P--wave $\pi$N partial waves $f_\pm^1(t)$ via 
$f_\pm^1(t) = F_\pi (t)\, J_\pm (t)$ in the conventional isospin 
decomposition, with the tabulated values of the $J_i (t)$ from \cite{LB}. For 
the pion charge form factor $F_\pi$ we use the standard Gounaris-Sakurai 
form \cite{GoSa}. 
We stress that the representation of Eq.~(\ref{uni}) gives the exact isovector 
spectral functions for $4M_\pi^2 \leq t \leq 16 M_\pi^2$ but in practice holds 
up to $t \simeq 50 M_\pi^2$. It has two distinct features. First, as already 
pointed out in \cite{FF}, it contains the important contribution of the $\rho$ 
meson (see Fig.~\ref{fig:2pic}) with its peak at $t \simeq 30 M_\pi^2$. 
Second, on the left shoulder of the $\rho$, the isovector spectral functions 
display a very pronounced enhancement close to the two--pion threshold. This 
is due to the logarithmic singularity on the second Riemann sheet located at 
$t_c = 4M_\pi^2 - M_\pi^4/m^2 = 3.98 M_\pi^2$, very close to the threshold. 
This pole comes from the projection of the nucleon Born graphs, or in modern 
language, from the triangle diagram also depicted in Fig.~\ref{fig:2pic}. If 
one were to neglect this important unitarity correction, one would severely 
underestimate the nucleon isovector radii \cite{HP2}. In fact, precisely the 
same effect is obtained at leading one--loop accuracy  in chiral perturbation 
theory, as 
discussed first in \cite{GSS,BKMrev}. This topic was further elaborated on in 
the framework of heavy baryon CHPT \cite{BKMspec,Norbert} and in a covariant 
calculation based on infrared regularization \cite{KM}. Thus, the most 
important two--pion contribution to the nucleon form factors can be determined 
by using either unitarity or CHPT (in the latter case, of course, the $\rho$ 
contribution is not included).
\medskip 

\noindent We now want to separate the (uncorrelated) pion contribution from 
the $\rho$--contribution in the isovector spectral functions, that is we 
decompose the isovector spectral functions as 
\beq 
\label{separation} 
{\rm Im}~G_I^V (t) = {\rm Im}~G_I^{V,2\pi} (t) + {\rm Im}~G_I^{V,\rho} (t)~, 
\quad I = E, M ~, 
\eeq 
and analogously for Im~$F_{1,2}^V (t)$. Using  Eq.~(\ref{dens}), we can then 
calculate the pion cloud contribution to the charge and magnetization density 
in the Breit--frame. The $\rho$--contribution in Eq.~(\ref{separation}) can be 
well represented by a Breit--Wigner form with a running width \cite{Norbert}, 
\beq
\label{rho} 
{\rm Im}~G^{V,\rho}_I (t) = \frac{b_I M_\rho^2 \sqrt{t} 
\Gamma_\rho (t)}{(M_\rho^2-t)^2 + t \Gamma_\rho^2 (t)}~, \quad I = E, M~, 
\eeq 
with the mass $M_\rho = 769.3\,$MeV and the width 
$\Gamma_\rho (t) = g^2(t-4M_\pi^2)^{3/2}/(48 \pi t)$, where the coupling 
$g = 6.03$ is determined from the empirical value 
$\Gamma_\rho (M_\rho^2) = 150.2\,$ MeV, and the parameters $b_I$ can be 
adjusted to the height of the resonance peak. The corresponding expressions 
for the imaginary parts of the Dirac and Pauli form factors can be obtained 
from Eq.~(\ref{conv}). It is clear that the separation into the (uncorrelated) 
pion contribution and the $\rho$--contribution introduces some 
model--dependence. To get an idea about the theoretical error induced by this 
procedure, we perform the separation in three different ways: 

\begin{itemize}
\item[(a)] The two--pion contribution can be directly obtained from the 
two--loop chiral perturbation theory calculation of \cite{Norbert}. Together 
with the $\rho$--contribution of Eq.~(\ref{rho}), this calculation gives a 
very good description of the empirical spectral functions.\footnote{Note that 
on the right side of the $\rho$, the two--loop representation is slightly 
larger than the empirical one, so that we expect to obtain an upper bound by 
employing  this procedure.} We will use the analytical formulae given in 
\cite{Norbert} where the low--energy constant $c_4$ was readjusted to 
avoid double counting of the $\rho$--contribution (see \cite{BKMlecs}). 
\item[(b)] A lower bound on the two--pion  contribution can be obtained by 
setting $F_\pi (t) = 1$ in  Eq.~(\ref{uni}). This prescription does not 
only remove the $\rho$--pole but also some small uncorrelated two--pion 
contributions contained in the pion form factor. 
\item[(c)] To obtain the two--pion contribution, we can also subtract 
Eq.~(\ref{rho}) from the spectral function Eq.~(\ref{uni}) including the full 
pion form factor.\footnote{A similar procedure was performed in \cite{UGMscal} 
to extract scalar meson properties from the scalar pion form factor.} The 
parameters $b_E=1.512$ and $b_M=5.114$ are determined such that the two--pion 
contribution at the $\rho$--resonance matches the two--loop chiral 
perturbation theory calculation of \cite{Norbert}. Variation of the $b_I$ 
around these values gives an additional error estimate. 
\end{itemize} 

\noindent Using these three methods, we obtain a fairly good handle on the 
theoretical accuracy of the non-resonant two--pion contribution. 

\medskip 

\noindent 
{\bf 4.} We have now collected all pieces to work out the density 
distribution of the two--pion contribution to the nucleon electromagnetic 
form factors. Before showing the results, some remarks are in order. As stated 
above, the spectral functions are determined by unitarity (or chiral 
perturbation theory) only up to some maximum value of $t$, denoted
$t_{\rm max}$ in the following. Thus, we have 
simply set the spectral functions in the integral Eq.~(\ref{dens}) to zero for 
momentum transfers beyond the value $t_{\rm max} = 40 M_\pi^2$. 
\begin{figure}[t] 
\parbox{.49\textwidth}{ 
\epsfig{file= 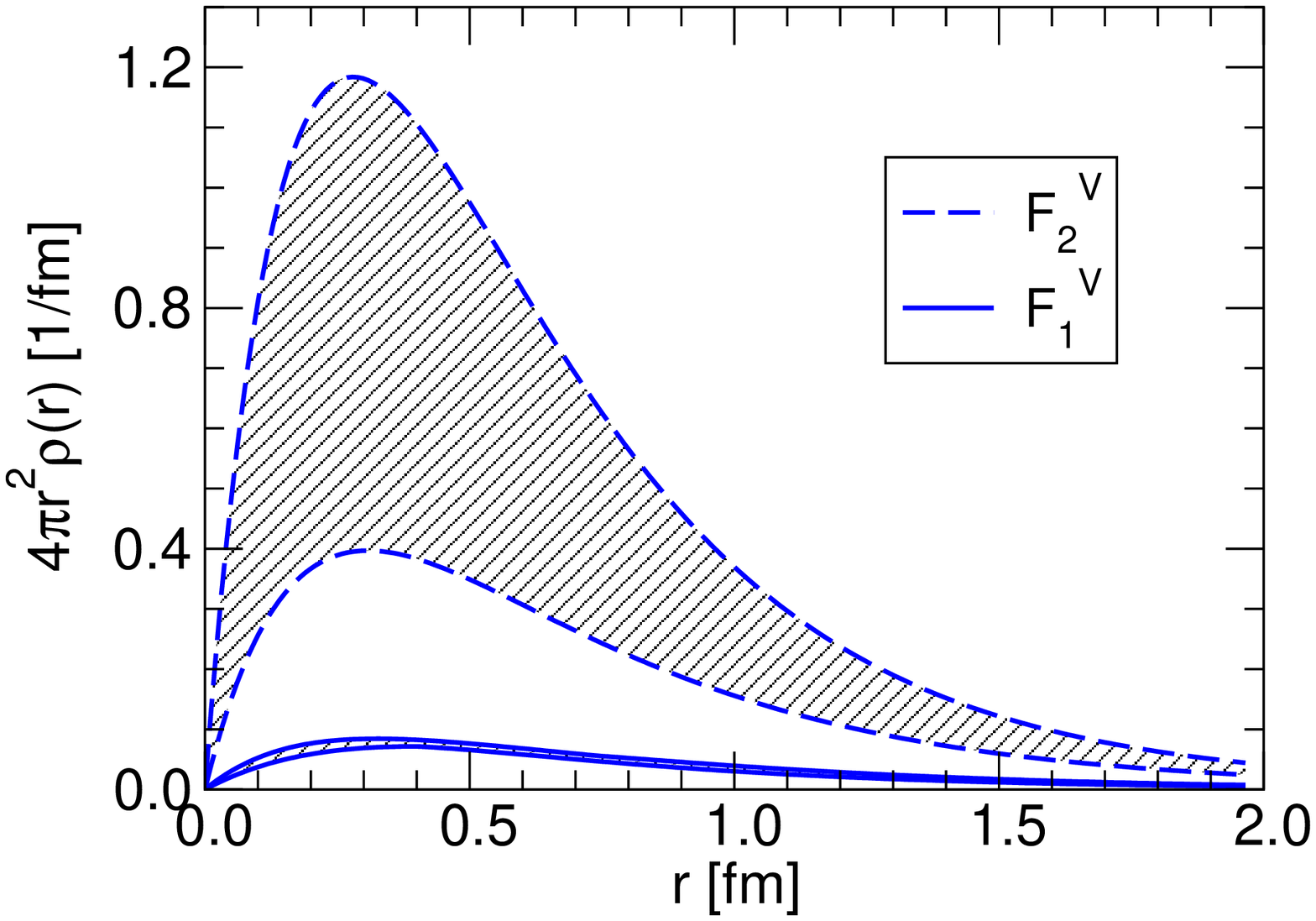,width=.48\textwidth,silent=,clip=}} 
\hfill \parbox{.49\textwidth}{ 
\epsfig{file= 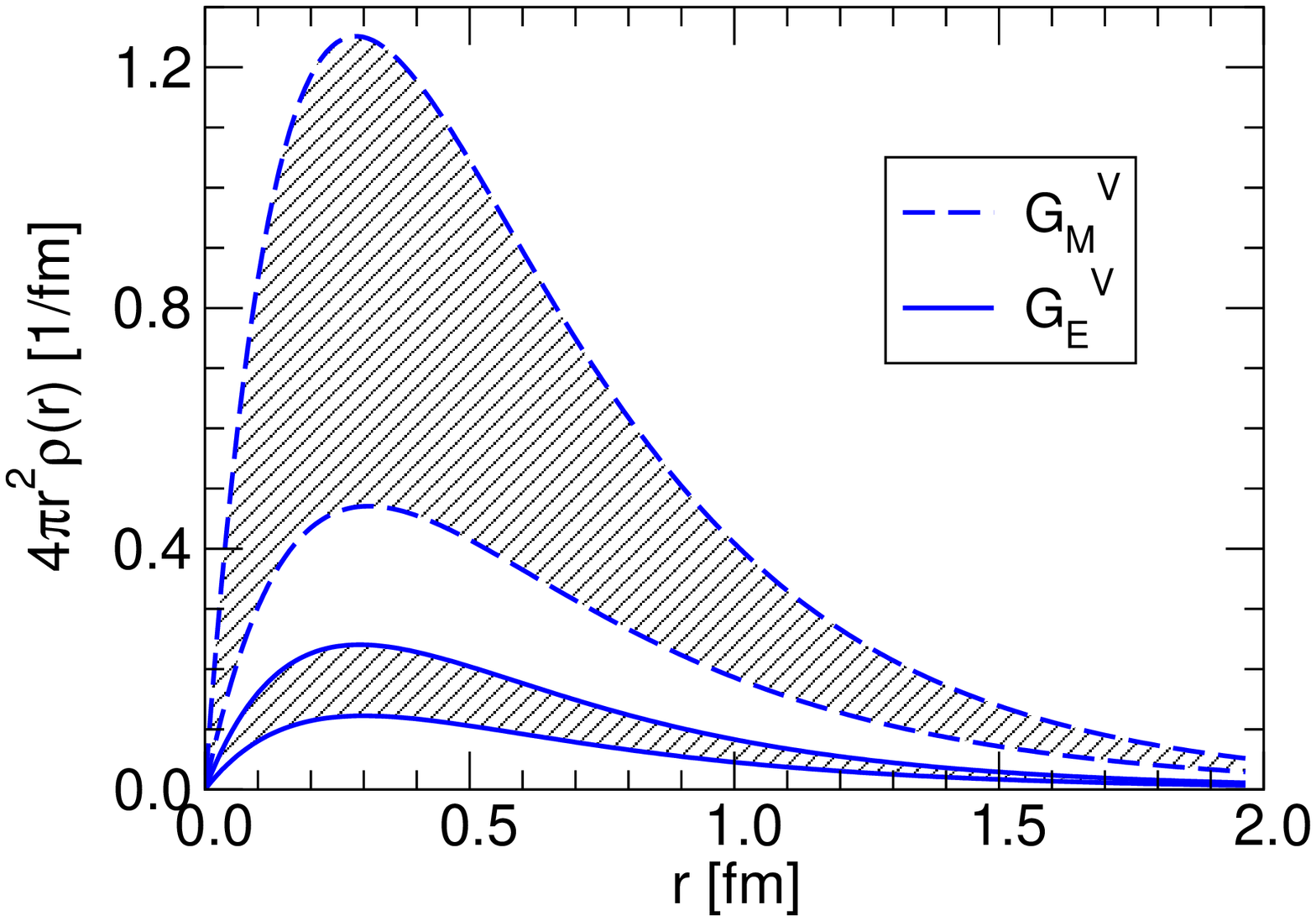,width=.48\textwidth,silent=,clip=}} 
\vspace{0.1cm} 
\begin{center} 
\caption{The densities of charge and magnetization due to the pion cloud. Left 
panel: $4\pi r^2 \rho(r)$ for the isovector Pauli (upper band) and Dirac 
(lower band) form factors. Right panel: $4\pi r^2 \rho(r)$ for the isovector 
magnetic (upper band) and electric (lower band) Sachs form factors. 
\label{fig:FF}} 
\end{center} 
\end{figure} 
In Fig.~\ref{fig:FF}, we show the resulting densities for the isovector form 
factors weighted with $4\pi r^2$. The contribution of the ``pion cloud'' to 
the total charge or magnetic moment is then simply obtained by integration 
over $r$. The bands reflect the theoretical uncertainty in the separation. For 
all form factors, the lower and upper bounds are given by methods (b) and (a), 
respectively. Method (c) generally yields a result between these bounds,
except for the Dirac form factor where it gives the upper bound. The weighted 
densities for the isovector Dirac and Pauli form factors are shown in the left 
panel of Fig.~\ref{fig:FF}. We see that these charge distributions show a 
pronounced peak around $r \simeq 0.3\,$fm, quite consistent with earlier 
determinations (see e.g. \cite{UGM,Holz}), and fall off smoothly with 
increasing distance. In the right panel of Fig.~\ref{fig:FF}, we show the 
densities (again weighted with $4\pi r^2$) for the electric and magnetic Sachs 
form factors which come out very similar to the case of the Dirac and  Pauli 
form factors. In comparison with Ref.~\cite{FW}, we generally obtain much 
smaller pion cloud effects at distances beyond 1~fm, e.g., 
by a factor 3 for $\rho_E^V (r)$ at 
$r=1.5$~fm.\footnote{Note that our results are not in disagreement with the 
recent Jefferson Lab data on the ratio of the proton electromagnetic form
factors \cite{Jones,Gayou}.
The effect observed there occurs at momentum transfers beyond 1 GeV$^2$
and is thus not related to the pion cloud.}

\medskip

\noindent
We have also studied the sensitivity of our results to the cut--off 
$t_{\rm{max}}$. While this may increase the value of the ``pion cloud'' 
contribution, it leaves the position of the maximum essentially unchanged. 
However, it is obvious from Eq.~(\ref{dens}) that masses beyond 1~GeV  and 
corresponding small--distance phenomena ($r\lesssim0.2$~fm) should not be 
related to the pion cloud of the nucleon. Finally, we show the corresponding 
two--pion contribution to the charges and radii for the various nucleon form 
factors in Table~\ref{radtab}. 
The contribution of the pion cloud to the isovector electric (magnetic) 
charge is 20\%~(10\%)
in the model of Ref.~\cite{FW}. This is consistent with our range of values 
for the electric charge but a factor of 1.5 smaller than our lower bound
for the magnetic one, see Table~\ref{radtab}.
Furthermore, note that the pion cloud gives only a fraction of all form factors
at zero momentum transfer. Normalized to the contribution of the 
pion cloud, the corresponding radii are of the order of 1~fm. In the model
of \cite{FW}, these radii are considerably larger, of the order of 1.5~fm.
Note that if one shifts all the strength of the corresponding spectral 
functions to threshold, one obtains an upper limit $r_{\rm max } = \sqrt{3/2}\,
M_\pi^{-1} \simeq 1.7\,$fm, assuming that the spectral functions are positive 
definite.
\renewcommand{\arraystretch}{1.3} 
\begin{table}[th] 
\begin{center} 
\begin{tabular}{|c|c|c|c||c|c|c|c|} 
\hline $F_1^V(0)$ & $F_2^V(0)$ & $G_E^V(0)$ & $G_M^V(0)$ & 
$\langle r^2\rangle_1^V$ & $\langle r^2\rangle_2^V$  & 
$\langle r^2\rangle_E^V$ & $\langle r^2\rangle_M^V$ \\ 
\hline\hline 
$0.07...0.08$ & $0.4...1.0$ & $0.1...0.2$ & $0.4...1.0$ & $0.1...0.2$ 
& $0.2...0.3$ & $0.2...0.3$ & $0.2...0.3$ \\ 
\hline 
\end{tabular} 
\end{center} 
\caption{\label{radtab} Two--pion contribution to charges and radii (in fm$^2$)
for the various nucleon form factors. The radii are normalized to the 
physical charges and magnetic moments.} 
\end{table} 

\medskip 

\noindent 
{\bf 5.} In this letter, we have considered the two--pion contribution to the 
nucleon isovector form factors. The corresponding spectral functions can be 
obtained from unitarity or chiral perturbation theory from threshold 
$t_0 = 4M_\pi^2$ up to $t_{\rm max} \simeq 40 M_\pi^2$ in a model--independent 
way. Subtracting the contribution from the dominant $\rho$--meson pole, we 
have constructed the uncorrelated two--pion component (loosely spoken the 
dominant effects of the nucleons' pion cloud). To obtain an estimate about 
the theoretical error of such a procedure, we have performed this subtraction 
in three different ways. The corresponding charge densities obtained by 
Fourier transforming the spectral functions peak at distances of about 0.3 fm 
and show no structure at the larger distances. 

\bigskip

\noindent
{\large {\bf Acknowledgements}}\\[0.1cm] 
We thank J.\ Friedrich and Th.\ Walcher as well as N.\ Kaiser
for useful discussions. This work 
was supported in part by the Deutsche Forschungsgemeinschaft (SFB 443). 
\vfill

\noindent 
 
\end{document}